\documentclass{article}
\usepackage{spconf,amsmath,amssymb,graphicx}
\usepackage{booktabs,array,siunitx}
\usepackage{url,cite,balance,capt-of}
\title{
INTERPRETING SELECTIVE-REVIEW COVERAGE\\
FOR HEAD-AND-NECK AUTO-CONTOURS
}

\name{
Jaden Moon$^{1,2}$
\qquad
Caiwen Jiang$^{2}$
\qquad
Songlin Zhao$^{2}$
\qquad
Wei Liu$^{2}$
}

\address{
$^{1}$Dartmouth College, Hanover, NH, USA \\
$^{2}$Mayo Clinic, Phoenix, AZ, USA
}

\newcommand{\ConfirmedEthicsStatement}{%
The source cohort and retrospective use of the associated clinical data were
approved by the Mayo Clinic Institutional Review Board under protocol
No.~24-011106. Given the retrospective design, the requirement for
study-specific written informed consent was waived under that protocol. The
secondary computational analysis reported here was approved under Mayo Clinic
IRB protocol No.~24-011064. The research was conducted in accordance with the
Declaration of Helsinki. The OpenKBP component used the open-access dataset
described by Babier et al.~\cite{babier2021openkbp}; its
dataset-construction study was approved by the University of Toronto Research
Ethics Board.}

\newcommand{\ConfirmedFundingStatement}{%
J.M. was supported by the Mayo Clinic Summer Undergraduate Research
Fellowship and the 2026 Passport AI Leaders Grant. No additional
external funding was received specifically for this work.}

\newcommand{\ConfirmedConflictStatement}{%
The authors declare no relevant financial or non-financial competing
interests.}

\newcommand{\ConfirmedCodeAvailabilityStatement}{%
The sanitized code and derived public-data artifacts needed to reproduce the
OpenKBP analyses will be released in a versioned public repository upon
acceptance. No Mayo Clinic data, identifiers, private model artifacts, or
restricted third-party assets will be included.}

\newcommand{\ConfirmedAIDisclosure}{%
ChatGPT and OpenAI Codex were used as assistive tools during software development and manuscript preparation. Codex generated or revised portions of the OpenKBP analysis, validation, figure-generation, and manuscript-checking scripts from author-specified endpoints, algorithms, thresholds, data splits, interfaces, and evaluation procedures. ChatGPT assisted with revising the abstract, Introduction, Materials and Methods, Results, and Discussion; LaTeX preparation; and the text and layouts of Figures 1–2 and Tables 1–3. The authors reviewed and modified all AI-assisted code, executed the analyses and checks, and verified the reported text, references, figures, and numerical results against retained outputs and source records. Neither ChatGPT nor Codex was invoked as part of the evaluated routing pipelines or used to generate study data, clinical labels, segmentation outputs, routing decisions, or reported numerical results. The authors take responsibility for the final manuscript and artifacts.}

\newcommand{\MayoRecordCount}{3968}
\newcommand{\MayoReferencePresentPairCount}{1004}
\newcommand{\MayoIncludedPairCount}{992}
\newcommand{\MayoExcludedPairCount}{12}

\newcommand{\MayoEmptyPredictionCount}{1167}
\newcommand{\MayoNonemptyPredictionCount}{2801}
\newcommand{\MayoEmptyPredictionPct}{29.41}
\newcommand{\MayoEndpointPositiveCount}{964}
\newcommand{\MayoEmptyEndpointPositiveCount}{0}
\newcommand{\MayoNonemptyEndpointNegativeCount}{1837}
\newcommand{\MayoContextualPassCount}{1524}
\newcommand{\MayoComparatorPassCount}{1534}
\newcommand{\MayoContextualCoveragePct}{38.41}
\newcommand{\MayoComparatorCoveragePct}{38.66}
\newcommand{\MayoContextualEmptyPassCount}{1167}
\newcommand{\MayoComparatorEmptyPassCount}{1167}
\newcommand{\MayoContextualEmptyPassSharePct}{76.57}
\newcommand{\MayoComparatorEmptyPassSharePct}{76.08}
\newcommand{\MayoContextualNonemptyPassCount}{357}
\newcommand{\MayoComparatorNonemptyPassCount}{367}
\newcommand{\MayoContextualNonemptyCoveragePct}{12.75}
\newcommand{\MayoComparatorNonemptyCoveragePct}{13.10}
\newcommand{\MayoContextualPositivePassCount}{3}
\newcommand{\MayoComparatorPositivePassCount}{9}
\newcommand{\MayoContextualEmptyPositivePassCount}{0}
\newcommand{\MayoComparatorEmptyPositivePassCount}{0}
\newcommand{\MayoContextualNonemptyNegativePassCount}{354}
\newcommand{\MayoComparatorNonemptyNegativePassCount}{358}

\newcommand{\MayoPositiveContextualOnlyCount}{3}
\newcommand{\MayoPositiveComparatorOnlyCount}{9}
\newcommand{\MayoPositiveBothReviewCount}{952}

\newcommand{\MayoRestrictedRecordCount}{3008}
\newcommand{\MayoRestrictedContextualPassCount}{564}
\newcommand{\MayoRestrictedComparatorPassCount}{574}
\newcommand{\MayoRestrictedContextualCoveragePct}{18.75}
\newcommand{\MayoRestrictedComparatorCoveragePct}{19.08}
\newcommand{\MayoReviewRuleContextualPassCount}{357}
\newcommand{\MayoReviewRuleComparatorPassCount}{367}
\newcommand{\MayoReviewRuleContextualCoveragePct}{9.00}
\newcommand{\MayoReviewRuleComparatorCoveragePct}{9.25}
\newcommand{\MayoResidualEmptyCount}{207}
\newcommand{\MayoResidualContextualPassSharePct}{36.70}
\newcommand{\MayoResidualComparatorPassSharePct}{36.06}
\newcommand{\MayoPlanEndpointCutoff}{28.4223}

\newcommand{\OpenKBPTestPatientCount}{100}
\newcommand{\OpenKBPTestCandidateCount}{5981}
\newcommand{\OpenKBPEndpointPositiveCount}{933}
\newcommand{\OpenKBPEvaluableCandidateCount}{5437}

\newcommand{\OpenKBPRouterCoveragePct}{26.38}
\newcommand{\OpenKBPRouterNonEvaluablePassCount}{544}
\newcommand{\OpenKBPEvaluablePassCount}{1034}
\newcommand{\OpenKBPRouterPositivePassCount}{12}
\newcommand{\OpenKBPRouterSensitivityPct}{98.71}

\newcommand{\OpenKBPGatedCoveragePct}{17.29}
\newcommand{\OpenKBPCoverageDiffPP}{-9.10}
\newcommand{\OpenKBPCoverageDiffLowerPP}{-9.15}
\newcommand{\OpenKBPCoverageDiffUpperPP}{-9.06}
\newcommand{\OpenKBPBootstrapCount}{10000}

\newcommand{\OpenKBPValidationGatePassCount}{493}
\newcommand{\OpenKBPTestAUROC}{0.833}
\newcommand{\OpenKBPTestAP}{0.449}
\newcommand{\OpenKBPTestPositivePrevalence}{0.156}
\newcommand{\OpenKBPLockedThreshold}{0.1218086865}
\newcommand{\OpenKBPExplicitEmptyControlCount}{542}
\newcommand{\OpenKBPErosionCollapseCount}{2}

\newcommand{\DevelopmentCandidateCount}{14223}

\newcommand{\DevelopmentEvaluableCount}{12926}
\newcommand{\DevelopmentNonEvaluableCount}{1297}

\newcommand{\DevelopmentBootstrapCount}{2000}
\newcommand{\OriginalRouterTotalCoveragePct}{23.43}
\newcommand{\EvaluableOnlyRouterTotalCoveragePct}{14.74}
\newcommand{\ThreeStateRouterTotalCoveragePct}{14.72}
\newcommand{\GatedRouterTotalCoveragePct}{14.31}
\newcommand{\GatedRouterEvaluableCoveragePct}{15.75}
\newcommand{\OriginalRouterEvaluableCoveragePct}{15.75}
\newcommand{\EvaluableOnlyRouterEvaluableCoveragePct}{16.22}
\newcommand{\ThreeStateRouterEvaluableCoveragePct}{16.19}
\newcommand{\OriginalRouterAUROC}{0.8265}
\newcommand{\OriginalRouterAP}{0.5270}
\newcommand{\OriginalRouterSensitivityPct}{99.05}
\newcommand{\EvaluableOnlyRouterAUROC}{0.8265}
\newcommand{\EvaluableOnlyRouterAP}{0.5287}
\newcommand{\EvaluableOnlyRouterSensitivityPct}{99.01}
\newcommand{\ThreeStateRouterAUROC}{0.8262}
\newcommand{\ThreeStateRouterAP}{0.5291}
\newcommand{\ThreeStateRouterSensitivityPct}{99.01}

\newcommand{\DevelopmentEmptyControlCount}{1290}
\newcommand{\DevelopmentErosionEmptyCount}{7}

\makeatletter
\def\ps@preprintfirst{%
  \def\@oddhead{}\def\@evenhead{}%
  \def\@oddfoot{\hfil\parbox[b]{0.90\textwidth}{\centering\fontsize{6}{7}\selectfont
  This work has been submitted to the IEEE for possible publication. Copyright may be transferred without notice, after which this version may no longer be accessible.}\hfil}%
  \let\@evenfoot\@oddfoot}
\makeatother

\begin{document}
\ninept
\maketitle
\thispagestyle{preprintfirst}

\begin{abstract}
Selective-review systems aim to reduce human review by passing candidates
predicted to be low risk. However, coverage---the fraction passed---can be
misleading when measurements needed to interpret risk are unavailable. We
separate structural evaluability (a nonempty mask with all required measurements
available), computational-endpoint status, and pass-or-review decisions using
archived Mayo records and controlled OpenKBP analyses. In
\num{\MayoRecordCount} Mayo records, all \num{\MayoEmptyPredictionCount}
recorded empty predictions---predicted target-organ masks containing zero voxels
despite nonempty reference masks---were endpoint-negative and passed both
modules, accounting for approximately 76\% of each pass set. In a held-out
OpenKBP test, a logistic-regression router achieved an area under the receiver
operating characteristic curve of \OpenKBPTestAUROC{} and average precision of
\OpenKBPTestAP{} but passed all
\num{\OpenKBPRouterNonEvaluablePassCount} non-evaluable candidates, including
\num{\OpenKBPExplicitEmptyControlCount} prespecified empty-mask controls and
\num{\OpenKBPErosionCollapseCount} erosions that became empty. A structural
gate removed those passes without changing scores or the threshold, preserved
all evaluable passes, and assigned \OpenKBPRouterSensitivityPct\% of
endpoint-positive candidates to review. Under the original preprocessing and in
patient-disjoint development evaluation, a router trained only on evaluable
candidates and a router modeling an unavailable state likewise passed no
non-evaluable candidates, with similar evaluable discrimination and at least
99.0\% of endpoint-positive candidates assigned to review in pooled outer-fold
predictions. These results show that coverage should be interpreted jointly with
candidate eligibility, endpoint scope, missingness handling, and routing rules;
they do not establish clinical validity, deployment safety, or natural-failure
prevalence.
\end{abstract}


\section{INTRODUCTION}
\label{sec:introduction}

Automated delineation of head-and-neck organs at risk (OARs) can reduce
contouring effort, but clinical acceptability remains a separate question
~\cite{nikolov2021clinical,lucido2023validation}. In selective-review
workflows, a routing model (router) assigns each candidate OAR mask to pass or
review. \emph{Coverage} is the pass fraction, and passed candidates form the
\emph{pass set}. A candidate is \emph{structurally evaluable} when its mask is
nonempty and all required candidate-derived measurements are available. The
\emph{computational endpoint} is the algorithmic label: endpoint-positive when
a predefined trigger activates and endpoint-negative otherwise. A pass is a
quality-assurance (QA) routing decision, not clinical
clearance~\cite{geifman2017selective}.

These quantities can disagree. An empty mask may make contour- and dose-derived
measurements unavailable. Under the endpoint rules examined here, unavailable
measurements did not activate an error trigger, and an uncomputable composite
received a negative label. A router could nevertheless assign a low-risk score
and pass the candidate. Coverage may therefore include candidates lacking the
evidence needed to interpret their endpoint.

Contour-QA methods predict segmentation quality, assess contour shape, or
combine geometric and dose tests
~\cite{luan2023hnqa,wooten2023shapeqa,poel2025duallayer,fankhauser2026catqa}.
Broader QA systems and studies of natural gross failures extend this line of
work~\cite{brooks2024knowledgeqa,temple2024grossfailure}. Our question is
complementary: whether candidates counted as passes have the evidence required
by the endpoint used to assign risk, consistent with problem-aware
validation~\cite{maierhein2024metricsreloaded}.

We combine archived Mayo pass-set characterization, a held-out OpenKBP test of
eligibility under a fixed router, and patient-disjoint comparisons of
alternative eligibility mechanisms. \textbf{Contributions.} We (1) provide a
reusable framework separating structural evaluability, endpoint status, and
routing decisions; (2) characterize non-evaluable passes in archived Mayo
records and show in a separate controlled test that endpoint discrimination
does not prevent such passes; and (3) compare structural gating with
evaluable-only and three-state routers while distinguishing population
restrictions from decision-rule changes. Figure~\ref{fig:coverage-framework}
summarizes the framework.

\section{MATERIALS AND METHODS}
\label{sec:methods}

Archived Mayo records describe fixed routing behavior, whereas controlled
OpenKBP analyses evaluate eligibility. OpenKBP preprocessing used training data,
thresholding used validation data, and the prespecified test was evaluated once;
remaining analyses used validation or development data.

\subsection{Data sources and evaluation units}
\label{subsec:cohort}

We analyzed archived single-center head-and-neck radiotherapy cases from Mayo
Clinic Arizona. Clinician reference OAR contours were consolidated into
15-label masks resized to $256\times256\times128$. Of
\num{\MayoReferencePresentPairCount} reference-present case--OAR pairs,
\num{\MayoIncludedPairCount} were included in the QA analysis and
\num{\MayoExcludedPairCount} were excluded by the original pipeline. Each
included pair contributed nnU-Net v2.8.0 predictions from fold 0 at epochs 50,
500, 950, and 1000, yielding \num{\MayoRecordCount} records
~\cite{isensee2021nnunet}.

These records did not constitute an independent patient-level validation set: case-group
identifiers were not verified to map one-to-one to patients, and four epoch
records came from each pair. The modules had been developed on \num{33260}
records from \num{676} identifiers. The contextual module used case-grouped
folds, whereas the comparator selected its threshold on the evaluation
partition. With models, labels, thresholds, and decisions fixed, the Mayo
analysis was descriptive.

A secondary population restriction excluded four structures selected by recorded
availability: bilateral cochleae, left eye, and optic chiasm (labels 2, 3, 5,
and 9). We
defined a recorded empty prediction as a predicted mask containing zero voxels
for the target OAR when the corresponding reference mask was nonempty; a
prediction was recorded nonempty when the target-OAR label contained at least
one voxel. Voxel counts linked the \num{\MayoEmptyPredictionCount} recorded
empty predictions to unavailable candidate dose summaries. These fields
establish the recorded structural state, not its cause; the original masks were
not re-inspected.

\subsection{Structural evaluability and computational endpoints}
\label{subsec:endpoints}

To separate eligibility from endpoint risk, let $Y_i=1$ denote endpoint
positivity and $A_i=1$ a pass. Both Mayo modules were evaluated against
$Y_i=y_i^{\mathrm{base}}\lor y_i^{\mathrm{plan}}$, combining the baseline-dose
flag and plan-derived endpoint; neither component was clinically adjudicated.
Endpoint-positive sensitivity is the fraction of endpoint-positive candidates
assigned to review, $1-\sum_iA_iY_i/\sum_iY_i$.

\medskip
\noindent\textbf{Baseline-dose flag.}
The baseline-dose flag compares candidate and reference mean dose, maximum dose,
and hottest-volume means (0.03, 0.1, and 1~cm$^3$). For finite
$d_{\mathrm{cand}},d_{\mathrm{ref}}$ with $|d_{\mathrm{ref}}|>10^{-8}$, it
activates if $|d_{\mathrm{cand}}-d_{\mathrm{ref}}|/|d_{\mathrm{ref}}|>0.2$.
It also activates when either fully finite summary sequence---candidate or
reference maximum dose followed by hottest-volume means for 0.03, 0.1, and
1~cm$^3$---contains a consecutive increase $>10^{-8}$. Missing differences do
not activate it.

\medskip
\noindent\textbf{Plan-derived endpoint.}
The plan-derived endpoint aggregates five families: dose-endpoint discrepancy,
local dose gradient, high-dose-shell involvement, shell proximity, and
candidate--reference disagreement. Training-only fitting retained 40 of 42
measurements. For measurement $x_{ij}$ of candidate $i$, the transformation
$z_{ij}=g_j(x_{ij})$ makes larger values indicate greater discrepancy. Let
$m_j,r_j$ be training centering and scaling constants; $q_{j,1},q_{j,99}$
clipping quantiles; $J_{if}$ the available measurements in family $f$; $F_i$
the available families; and $q_{\mathrm{endpoint}}$ the endpoint cutoff. With
$[a]_+=\max(a,0)$ and $\mathbf{1}\{\cdot\}$ denoting the indicator,
\begin{equation}
\begin{aligned}
c_{ij}&=\left[\frac{\operatorname{clip}(z_{ij},q_{j,1},q_{j,99})-m_j}{r_j}\right]_+,\\
\bar c_{if}&=\frac{\sum_{j\in J_{if}}c_{ij}}{|J_{if}|},\quad
s_i=\max_{f\in F_i}\bar c_{if},\\
y_i^{\mathrm{plan}}&=\mathbf{1}\{\operatorname{finite}(s_i)\land s_i\ge q_{\mathrm{endpoint}}\}.
\end{aligned}
\label{eq:plan-endpoint}
\end{equation}
Discrepancies are clipped, standardized, floored at zero, averaged within
families, and combined by the largest family mean. The cutoff was the
training 97.5th percentile, $q_{\mathrm{endpoint}}\approx\MayoPlanEndpointCutoff$.
Without a computable family mean, the label remains negative. Thus,
endpoint-negative can indicate low discrepancy or insufficient evidence.

\subsection{Fixed routing modules}
\label{subsec:routers}

We analyzed the fixed modules that generated the archived pass sets. The \textbf{comparator} predicts $y_i^{\mathrm{base}}$ with a random
forest using 91 metadata, geometry, and dose inputs. Candidates with finite
scores at or below $0.13767$ passed; all others were assigned to review.
Four-epoch summaries indicated whether values were finite, whereas direct
contour features remained numeric. Because the saved training matrices were unavailable, we analyzed saved outputs.

The \textbf{contextual module} predicts $Y_i$ with histogram-gradient boosting
from 65 inputs: 36 metadata variables, five current-candidate dose summaries,
20 summaries across four epochs, and four availability indicators. Candidates
with finite scores below $0.04529$ passed; all others were assigned to review.
The model did not meet its prespecified coverage or coverage-improvement
criteria. The cutoffs and comparison defining improvement were not recoverable,
so we report it descriptively.

Neither module enforced mandatory eligibility. Because they differ in target,
inputs, estimator, preprocessing, and threshold selection, their pass counts
describe archived behavior, not controlled comparative performance.

\subsection{Record linkage and coverage definitions}
\label{subsec:structural}
\label{subsec:metrics}

To characterize the structural states counted as passes, we linked prediction
voxel counts to dose-summary availability. Let $R_i$ and $P_i$ be reference and
prediction voxel counts, with all $R_i>0$. Thus, $E_i=1$ identifies a recorded
empty prediction and $H_i=1$ a recorded nonempty prediction:
\[
E_i=\mathbf{1}\{P_i=0\},\qquad H_i=\mathbf{1}\{P_i>0\}.
\]
We compared $H_i$ with five candidate dose-summary availability fields. Joins
by case identifier, OAR label, and epoch yielded no duplicate, missing, or
mismatched records. The linkage identifies pass-set composition, not upstream
cause.

We then compared coverage in the original population, after excluding four
structures, among recorded nonempty predictions, and under an empty-to-review
rule. Let $N$ denote the original record count and $L_i=1$ identify the four
selected structures:
\begin{equation}
\begin{aligned}
C&=\frac{\sum_i A_i}{N},&
C_{\mathrm{excl}}&=\frac{\sum_i A_i(1-L_i)}{\sum_i(1-L_i)},\\
C_{\mathrm{nonempty}}&=\frac{\sum_i A_iH_i}{\sum_i H_i},&
C_{\mathrm{gate}}&=\frac{\sum_i A_iH_i}{N}.
\end{aligned}
\label{eq:coverage-denominators}
\end{equation}
Exclusion and nonempty-only analyses change the population while preserving
decisions. The empty-to-review rule retains the population and changes only
routing assignments. None of these analyses retrains a model or evaluates a deployed workflow.

\begin{figure*}[t]
  \centering
  \begin{minipage}[t]{0.485\textwidth}
    \vspace{0pt}
    \centering
    \includegraphics[width=\linewidth]
    {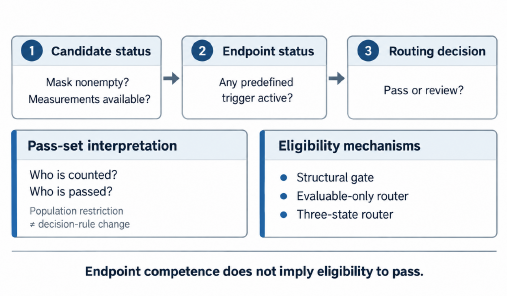}
    \captionof{figure}{Structural evaluability, computational-endpoint status,
    and routing decisions are analyzed separately to interpret coverage.
    The endpoint is an evaluation label, not an input to the routing decision.
    Population restrictions change who is counted; structural gating and
    learned alternatives change who can pass.}
    \label{fig:coverage-framework}
  \end{minipage}
  \hfill
  \begin{minipage}[t]{0.485\textwidth}
    \vspace{0pt}
    \centering
    \includegraphics[width=\linewidth]
    {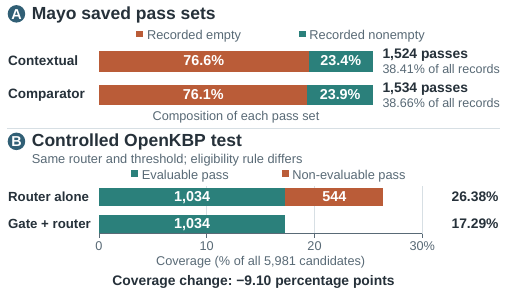}
    \captionof{figure}{Pass-set composition in the archived Mayo records and
    controlled OpenKBP test. (A) Recorded empty predictions constituted about
    76\% of both Mayo pass sets, although they represented 29.41\% of all
    archived records. (B) With the router and threshold unchanged, the
    structural gate removed all 544 non-evaluable passes---542 prespecified
    empty-mask controls and two erosions that became empty---while retaining
    all 1,034 evaluable passes.}
    \label{fig:pass-composition}
  \end{minipage}
\end{figure*}

\subsection{Controlled OpenKBP experiment}
\label{subsec:openkbp-methods}

To evaluate eligibility under controlled conditions, we used the official
OpenKBP split: 200 training, 40 validation, and 100 test
patients~\cite{babier2021openkbp}. Each row represented one candidate mask for a
patient--OAR pair. For seven OARs, candidates included unchanged reference
masks, translations, erosions, dilations, component deletions, truncations, and
prespecified empty-mask controls, yielding 2,407 validation and 5,981 test rows.

OpenKBP used the fixed baseline-dose flag, $Y_i=y_i^{\mathrm{base}}$
(Section~\ref{subsec:endpoints}), not the Mayo plan-derived composite.
Candidate--reference dose comparisons defined labels but were excluded from
router and gate inputs. An $\ell_2$-penalized logistic-regression router with
balanced class weights used candidate geometry, dose, OAR identity, and
availability with training-only preprocessing. Candidates with finite scores
below the threshold passed; all others were assigned to review. We selected the
maximum-coverage validation threshold at empirical endpoint-positive
sensitivity $\ge0.99$ and fixed it at \OpenKBPLockedThreshold{} for the single
test evaluation.

To isolate eligibility from endpoint scoring, we applied
$A_i^{\mathrm{gate}}=A_iS_i$ without changing scores or the threshold; $S_i=1$
requires a nonempty candidate with all required summaries computable. We
computed held-out AUROC, AP, and pass composition; \num{\OpenKBPBootstrapCount}
patient-cluster bootstrap resamples quantified uncertainty.

Validation-only controls examined threshold tightening, perturbation severity,
and all 31 nonempty missingness patterns in 218 unchanged reference masks.
Separately, a rule assigning candidates with unavailable required evidence to
review matched gating on saved validation and test decisions.

\subsection{Eligibility-aware development comparison}
\label{subsec:eligibility-models}

In 240 OpenKBP development patients, we compared four models: Model~A, a binary
baseline refit within each outer fold; Model~B, fit only on evaluable rows;
Model~C, a three-state model (negative, positive, or unavailable) with risk
$1-P(\mathrm{endpoint\mbox{-}negative})$; and Model~D, Model~A followed by the
gate.

Models B and C excluded evaluability, nonemptiness, availability, and missingness
indicators. Models A--C used $\ell_2$ regularization ($C=1$) without class
weighting; Models A and B used binary logistic regression with liblinear and Model~C
used multinomial logistic regression with lbfgs. Training-fold preprocessing
median-imputed and standardized numerical features and one-hot encoded OAR
identity. Model~B was thresholded on evaluable rows and scored on all candidates.

To avoid patient overlap, five outer folds held out patients and four inner folds
selected maximum-coverage thresholds at empirical endpoint-positive sensitivity
$\ge0.99$. We computed metrics from pooled outer-fold predictions and quantified
uncertainty with \num{\DevelopmentBootstrapCount} patient-cluster bootstrap
resamples. Total
coverage used all \num{\DevelopmentCandidateCount} candidates; evaluable
coverage and AUROC/AP used the \num{\DevelopmentEvaluableCount} structurally
evaluable candidates; sensitivity used endpoint-positive candidates. No test
rows entered this development-only comparison. Additional development analyses
tested representation and OAR dependence by varying imputation with models and
thresholds fixed and by excluding each OAR from fitting and threshold selection.

\section{RESULTS}
\label{sec:results}

\subsection{Recorded empty predictions dominated pass sets}
\label{subsec:structural-results}

Of \num{\MayoRecordCount} archived records,
\num{\MayoEmptyPredictionCount} (\MayoEmptyPredictionPct\%) had a recorded
empty prediction and a nonempty reference. Under the endpoint rules,
unavailable measurements did not activate an error trigger, so all were
endpoint-negative. Both modules passed every such record; these accounted for
\MayoContextualEmptyPassSharePct\% and
\MayoComparatorEmptyPassSharePct\% of contextual and comparator passes,
respectively (Fig.~\ref{fig:pass-composition}A). All
\num{\MayoEndpointPositiveCount} endpoint-positive records were recorded
nonempty (Table~\ref{tab:structural-routing}).

After excluding the four availability-selected structures,
\num{\MayoResidualEmptyCount} recorded empty predictions remained and formed
\MayoResidualContextualPassSharePct\% and
\MayoResidualComparatorPassSharePct\% of the pass sets. Recorded empty predictions therefore persisted beyond those structures. Voxel
counts and availability fields agreed, and linkage checks found no
discrepancies; the original masks were not re-inspected, so the upstream cause
remains unresolved.

\begin{table}[ht]
\centering
\caption{Pass counts by structural status and computational endpoint. Labels
are algorithmic, not clinical adjudications.}
\label{tab:structural-routing}
\footnotesize
\setlength{\tabcolsep}{2.5pt}
\renewcommand{\arraystretch}{1.00}
\begin{tabular*}{\columnwidth}{@{\extracolsep{\fill}}lrrr@{}}
\toprule
\shortstack[l]{Structural status /\\endpoint} &
Records &
\shortstack{Contextual\\passes} &
\shortstack{Comparator\\passes} \\
\midrule
\shortstack[l]{Recorded empty /\\endpoint-negative} &
\num{\MayoEmptyPredictionCount} &
\num{\MayoContextualEmptyPassCount} &
\num{\MayoComparatorEmptyPassCount} \\
\shortstack[l]{Recorded empty /\\endpoint-positive} &
\num{\MayoEmptyEndpointPositiveCount} &
\num{\MayoContextualEmptyPositivePassCount} &
\num{\MayoComparatorEmptyPositivePassCount} \\
\shortstack[l]{Recorded nonempty /\\endpoint-negative} &
\num{\MayoNonemptyEndpointNegativeCount} &
\num{\MayoContextualNonemptyNegativePassCount} &
\num{\MayoComparatorNonemptyNegativePassCount} \\
\shortstack[l]{Recorded nonempty /\\endpoint-positive} &
\num{\MayoEndpointPositiveCount} &
\num{\MayoContextualPositivePassCount} &
\num{\MayoComparatorPositivePassCount} \\
\midrule
\textbf{Recorded nonempty total} &
\textbf{\num{\MayoNonemptyPredictionCount}} &
\textbf{\num{\MayoContextualNonemptyPassCount}} &
\textbf{\num{\MayoComparatorNonemptyPassCount}} \\
\textbf{All records} &
\textbf{\num{\MayoRecordCount}} &
\textbf{\num{\MayoContextualPassCount}} &
\textbf{\num{\MayoComparatorPassCount}} \\
\bottomrule
\end{tabular*}
\end{table}

\subsection{Similar nonempty coverage masked decision differences}
\label{subsec:nonempty-results}

Among the \num{\MayoNonemptyPredictionCount} recorded nonempty predictions in
the original Mayo cohort, the contextual and comparator modules passed
\num{\MayoContextualNonemptyPassCount}
(\MayoContextualNonemptyCoveragePct\%) and
\num{\MayoComparatorNonemptyPassCount}
(\MayoComparatorNonemptyCoveragePct\%), respectively. The modules passed different endpoint-positive records:
\num{\MayoPositiveContextualOnlyCount} contextual only,
\num{\MayoPositiveComparatorOnlyCount} comparator only, none by both, and
\num{\MayoPositiveBothReviewCount} by neither. Thus, similar coverage described pass-set size, not agreement. Because the modules were not controlled comparators, the counts do not rank
clinical performance.

\subsection{Population restrictions and gating changed different quantities}
\label{subsec:coverage-results}

We compared population restrictions with the empty-to-review rule. Restrictions
changed the population but not saved decisions; the rule retained all
\num{\MayoRecordCount} records and changed only routing assignments. It reduced coverage from
\MayoContextualCoveragePct\% to \MayoReviewRuleContextualCoveragePct\% and from
\MayoComparatorCoveragePct\% to \MayoReviewRuleComparatorCoveragePct\%
(Table~\ref{tab:coverage-definitions}). No model, endpoint, score, or threshold
changed.

\par\noindent
\begin{minipage}{\columnwidth}
\captionof{table}{Coverage under population restrictions and an empty-to-review
rule. Restrictions preserve decisions; the rule changes them.}
\label{tab:coverage-definitions}
\centering
\scriptsize
\setlength{\tabcolsep}{2.4pt}
\renewcommand{\arraystretch}{0.90}
\begin{tabular*}{\columnwidth}{@{\extracolsep{\fill}}lrr@{}}
\toprule
Definition & Contextual & Comparator \\
\midrule
Original population &
\shortstack[r]{\num{\MayoContextualPassCount}/\num{\MayoRecordCount}\\(\MayoContextualCoveragePct\%)} &
\shortstack[r]{\num{\MayoComparatorPassCount}/\num{\MayoRecordCount}\\(\MayoComparatorCoveragePct\%)} \\
Four structures excluded &
\shortstack[r]{\num{\MayoRestrictedContextualPassCount}/\num{\MayoRestrictedRecordCount}\\(\MayoRestrictedContextualCoveragePct\%)} &
\shortstack[r]{\num{\MayoRestrictedComparatorPassCount}/\num{\MayoRestrictedRecordCount}\\(\MayoRestrictedComparatorCoveragePct\%)} \\
Recorded nonempty only &
\shortstack[r]{\num{\MayoContextualNonemptyPassCount}/\num{\MayoNonemptyPredictionCount}\\(\MayoContextualNonemptyCoveragePct\%)} &
\shortstack[r]{\num{\MayoComparatorNonemptyPassCount}/\num{\MayoNonemptyPredictionCount}\\(\MayoComparatorNonemptyCoveragePct\%)} \\
Empty-to-review rule &
\shortstack[r]{\num{\MayoReviewRuleContextualPassCount}/\num{\MayoRecordCount}\\(\MayoReviewRuleContextualCoveragePct\%)} &
\shortstack[r]{\num{\MayoReviewRuleComparatorPassCount}/\num{\MayoRecordCount}\\(\MayoReviewRuleComparatorCoveragePct\%)} \\
\bottomrule
\end{tabular*}
\end{minipage}
\vspace{-2pt}

\subsection{Learned alternatives passed no non-evaluable candidates}
\label{subsec:eligibility-model-results}

Among \num{\DevelopmentCandidateCount} development candidates,
\num{\DevelopmentNonEvaluableCount} were non-evaluable:
\num{\DevelopmentEmptyControlCount} prespecified empty-mask controls and
\num{\DevelopmentErosionEmptyCount} erosions that became empty. Model~A passed
all; Models B and C passed none in any outer fold despite no explicit
evaluability, nonemptiness, availability, or missingness indicators.

\begin{center}
\captionof{table}{Pooled development results (no test rows). Denominators:
\num{\DevelopmentCandidateCount} total, \num{\DevelopmentEvaluableCount}
evaluable, and \num{2312} endpoint-positive. A: baseline; B: evaluable-only;
C: three-state; D: A plus gate.}
\label{tab:eligibility-models}
\footnotesize
\setlength{\tabcolsep}{1.5pt}
\renewcommand{\arraystretch}{0.90}
\begin{tabular*}{\columnwidth}{@{\extracolsep{\fill}}lrrrrr@{}}
\toprule
Model & \shortstack{Total\\cov.} & \shortstack{Evaluable\\cov.} &
\shortstack{Non-evaluable\\passes, $n$} & \shortstack{Evaluable\\AUROC/AP} &
\shortstack{Endpoint-\\positive\\sensitivity} \\
\midrule
A & \OriginalRouterTotalCoveragePct\% & \OriginalRouterEvaluableCoveragePct\% &
\num{1297} & \OriginalRouterAUROC/\OriginalRouterAP & \OriginalRouterSensitivityPct\% \\
B & \EvaluableOnlyRouterTotalCoveragePct\% & \EvaluableOnlyRouterEvaluableCoveragePct\% &
0 & \EvaluableOnlyRouterAUROC/\EvaluableOnlyRouterAP & \EvaluableOnlyRouterSensitivityPct\% \\
C & \ThreeStateRouterTotalCoveragePct\% & \ThreeStateRouterEvaluableCoveragePct\% &
0 & \ThreeStateRouterAUROC/\ThreeStateRouterAP & \ThreeStateRouterSensitivityPct\% \\
D & \GatedRouterTotalCoveragePct\% & \GatedRouterEvaluableCoveragePct\% &
0 & \OriginalRouterAUROC/\OriginalRouterAP & \OriginalRouterSensitivityPct\% \\
\bottomrule
\end{tabular*}
\end{center}
\vspace{-2pt}

Compared with Model~A, Models B and C had similar pooled evaluable coverage,
AUROC/AP, and endpoint-positive sensitivity, but lower total coverage
(Table~\ref{tab:eligibility-models}). Bootstrap intervals showed negligible
AUROC and sensitivity changes and a small AP increase for Model~B. Model~D
retained Model~A's evaluable metrics and passed no non-evaluable candidates.
Additional analyses showed that Model~B's behavior varied with imputation and
both learned alternatives had lower sensitivity on OARs excluded from fitting. Table~\ref{tab:eligibility-models}
therefore applies only to the original preprocessing and development
distribution.

\subsection{Endpoint discrimination did not ensure evaluability}
\label{subsec:openkbp-results}

Across all \num{\OpenKBPTestCandidateCount} candidates from
\num{\OpenKBPTestPatientCount} test patients, the fixed OpenKBP router achieved
computational-endpoint AUROC \OpenKBPTestAUROC{} and AP \OpenKBPTestAP{} at
endpoint-positive prevalence \OpenKBPTestPositivePrevalence. Nevertheless, it passed all
\num{\OpenKBPRouterNonEvaluablePassCount} non-evaluable candidates, including
\num{\OpenKBPExplicitEmptyControlCount} empty-mask controls and
\num{\OpenKBPErosionCollapseCount} erosions that became empty.

With scores and threshold unchanged, gating reduced total coverage from
\OpenKBPRouterCoveragePct\% to \OpenKBPGatedCoveragePct\% (change,
\OpenKBPCoverageDiffPP{} percentage points; 95\% confidence interval,
\OpenKBPCoverageDiffLowerPP{} to \OpenKBPCoverageDiffUpperPP). Both policies
passed \num{\OpenKBPRouterPositivePassCount} of
\num{\OpenKBPEndpointPositiveCount} endpoint-positive candidates and the same
\num{\OpenKBPEvaluablePassCount} of \num{\OpenKBPEvaluableCandidateCount}
evaluable candidates, preserving \OpenKBPRouterSensitivityPct\% pooled
endpoint-positive sensitivity. Thus, the gate removed only non-evaluable passes
(Fig.~\ref{fig:pass-composition}B).

Validation thresholding could not reproduce the gate: no score threshold
eliminated all 218 non-evaluable candidates while retaining any passes. At the same \num{\OpenKBPValidationGatePassCount}-pass
total, thresholding retained 275 evaluable and 218 non-evaluable candidates,
whereas gating retained 493 evaluable candidates and no non-evaluable
candidates. Compared with 2-mm perturbations, 8-mm translations and 4-mm
erosions/dilations increased endpoint positivity by 21--27 percentage points;
each dose-summary trigger also activated more often. Masking all five summaries
in 218 unchanged, nonempty validation reference masks left 133 passes (61.0\%;
95\% confidence interval, 56.8--65.1\%); gating assigned all 218 to review. The
direct unavailable-to-review rule matched the gate on saved validation and test
decisions without using Model~C.

\section{DISCUSSION}
\label{sec:discussion}

A router can discriminate its computational endpoint yet pass candidates that
lack the evidence required to interpret that endpoint. Prior contour-QA and
natural-failure studies evaluate contour quality or identify failures; our
results establish a complementary prerequisite for selective review: the
evidence required by the endpoint must be available before an automatic pass is
interpretable~\cite{luan2023hnqa,wooten2023shapeqa,poel2025duallayer,
brooks2024knowledgeqa,temple2024grossfailure}. Coverage should therefore be
reported with the denominator, pass-set composition, missingness handling,
endpoint definition and discrimination, eligibility rule, and routing rule.

The Mayo archive illustrates this mismatch. Recorded empty predictions lacked
candidate dose summaries, remained endpoint-negative, and passed both modules.
After excluding four selected structures,
\num{\MayoResidualEmptyCount} such records still formed more than 36\% of each
pass set, while similar coverage concealed different endpoint-positive
decisions. Because we did not re-inspect the original masks, these records establish
archived pass-set composition rather than the cause of the recorded empty states.

The controlled OpenKBP test isolated the same distinction under fixed router
scores and threshold. Although the router achieved useful held-out
discrimination, it passed every non-evaluable candidate; threshold tightening
removed evaluable candidates first because the non-evaluable candidates
occupied the low-risk tail. Controlled perturbations activated the endpoint
more often as their severity increased, whereas masking required summaries
still allowed some nonempty candidates to pass. The structural gate removed
non-evaluable passes while preserving evaluable decisions. In development data,
evaluable-only and three-state routers also passed no non-evaluable candidates
under the original preprocessing. However, Model B was imputation-dependent, and both learned alternatives had
lower sensitivity on OARs excluded from fitting, limiting this evidence to the
controlled development distribution.

Both data sources limit generalization. In the Mayo archive, identifiers may
not map uniquely to patients, each case--OAR pair contributed repeated
epoch-specific records, the comparator threshold was selected on the evaluation
partition, the endpoints lacked clinical adjudication, and the original masks
were not re-inspected. In OpenKBP, 542 of 544 held-out non-evaluable candidates
were prespecified empty-mask controls, and no natural-output or clinical
workflow study was performed. Future work should evaluate outputs from fixed
segmentation models using clinician-adjudicated endpoints and prospective
workflows.

Overall, endpoint risk and eligibility for automatic passage are distinct.
Selective-review reports should state the eligibility rule, denominator,
missingness handling, endpoint scope, and routing decision. These findings do not establish clinical validity, deployment safety,
workflow benefit, or natural-failure prevalence.

\newpage
\nobalance

\section{Compliance with Ethical Standards}
\label{sec:ethics}

\ConfirmedEthicsStatement

\section{Acknowledgments}
\label{sec:acknowledgments}

\paragraph*{Funding.}
\ConfirmedFundingStatement

\paragraph*{Competing interests.}
\ConfirmedConflictStatement

\paragraph*{Code availability.}
\ConfirmedCodeAvailabilityStatement

\paragraph*{Artificial-intelligence assistance.}
\ConfirmedAIDisclosure

\bibliographystyle{IEEEbib}
\bibliography{strings,refs}

@article{nikolov2021clinical,
  author  = {Nikolov, Stanislav and Blackwell, Sam and Zverovitch, Alexei and Mendes, Ruheena and Livne, Michelle and De Fauw, Jeffrey and Patel, Yojan and Meyer, Clemens and Askham, Harry and Romera-Paredes, Bernadino and Kelly, Christopher and Karthikesalingam, Alan and Chu, Carlton and Carnell, Dawn and Boon, Cheng and D'Souza, Derek and Moinuddin, Syed Ali and Garie, Bethany and McQuinlan, Yasmin and Ireland, Sarah and Hampton, Kiarna and Fuller, Krystle and Montgomery, Hugh and Rees, Geraint and Suleyman, Mustafa and Back, Trevor and Hughes, C{\'i}an Owen and Ledsam, Joseph R. and Ronneberger, Olaf},
  title   = {Clinically Applicable Segmentation of Head and Neck Anatomy for Radiotherapy: Deep Learning Algorithm Development and Validation Study},
  journal = {Journal of Medical Internet Research},
  year    = {2021},
  volume  = {23},
  number  = {7},
  pages   = {e26151},
  doi     = {10.2196/26151}
}

@article{lucido2023validation,
  author  = {Lucido, J. John and DeWees, Todd A. and Leavitt, Todd R. and Anand, Aman and Beltran, Chris J. and Brooke, Mark D. and Buroker, Justine R. and Foote, Robert L. and Foss, Olivia R. and Gleason, Angela M. and Hodge, Teresa L. and Hughes, C{\'i}an O. and Hunzeker, Ashley E. and Laack, Nadia N. and Lenz, Tamra K. and Livne, Michelle and Morigami, Megumi and Moseley, Douglas J. and Undahl, Lisa M. and Patel, Yojan and Tryggestad, Erik J. and Walker, Megan Z. and Zverovitch, Alexei and Patel, Samir H.},
  title   = {Validation of clinical acceptability of deep-learning-based automated segmentation of organs-at-risk for head-and-neck radiotherapy treatment planning},
  journal = {Frontiers in Oncology},
  year    = {2023},
  volume  = {13},
  pages   = {1137803},
  doi     = {10.3389/fonc.2023.1137803}
}

@inproceedings{geifman2017selective,
  author    = {Geifman, Yonatan and El-Yaniv, Ran},
  title     = {Selective Classification for Deep Neural Networks},
  booktitle = {Advances in Neural Information Processing Systems},
  year      = {2017},
  volume    = {30},
  pages     = {4878--4887},
  publisher = {Curran Associates, Inc.}
}

@article{luan2023hnqa,
  author  = {Luan, Shunyao and Xue, Xudong and Wei, Changchao and Ding, Yi and Zhu, Benpeng and Wei, Wei},
  title   = {Machine Learning-Based Quality Assurance for Automatic Segmentation of Head-and-Neck Organs-at-Risk in Radiotherapy},
  journal = {Technology in Cancer Research \& Treatment},
  year    = {2023},
  volume  = {22},
  pages   = {15330338231157936},
  doi     = {10.1177/15330338231157936}
}

@inproceedings{wooten2023shapeqa,
  author    = {Wooten, Zachary T. and Yu, Cenji and Court, Laurence E. and Peterson, Christine B.},
  title     = {Predictive Modeling Using Shape Statistics for Interpretable and Robust Quality Assurance of Automated Contours in Radiation Treatment Planning},
  booktitle = {Pacific Symposium on Biocomputing 2023},
  year      = {2023},
  volume    = {28},
  pages     = {395--406},
  doi       = {10.1142/9789811270611_0036}
}

@article{poel2025duallayer,
  author  = {Poel, Robert and Kamath, Amith and Ermi{\c{s}}, Ekin and Willmann, Jonas and R{\"u}fenacht, Elias and Andratschke, Nicolaus and Manser, Peter and Aebersold, Daniel M. and Reyes, Mauricio},
  title   = {A dual-layer quality assurance approach leveraging dose prediction for efficient review of automated contours of organs at risk in the brain in radiotherapy},
  journal = {Physics and Imaging in Radiation Oncology},
  year    = {2025},
  volume  = {36},
  pages   = {100888},
  doi     = {10.1016/j.phro.2025.100888}
}

@article{fankhauser2026catqa,
  author  = {Fankhauser, Lisa Stefanie and Toro, Maria Giulia and Smolders, Andreas Johan and Bellotti, Renato and Lomax, Antony John and Albertini, Francesca},
  title   = {Contour Assessment Tool for Quality Assurance ({CAT-QA}) to Speed Up Online Adaptive Radiotherapy},
  journal = {Physics in Medicine \& Biology},
  year    = {2026},
  volume  = {71},
  number  = {2},
  pages   = {025022},
  doi     = {10.1088/1361-6560/ae39a0}
}

@article{brooks2024knowledgeqa,
  author  = {Brooks, Jamison and Tryggestad, Erik and Anand, Aman and Beltran, Chris and Foote, Robert and Lucido, J. John and Laack, Nadia N. and Routman, David and Patel, Samir H. and Seetamsetty, Srinivas and Moseley, Douglas},
  title   = {Knowledge-based quality assurance of a comprehensive set of organ at risk contours for head and neck radiotherapy},
  journal = {Frontiers in Oncology},
  year    = {2024},
  volume  = {14},
  pages   = {1295251},
  doi     = {10.3389/fonc.2024.1295251}
}

@article{temple2024grossfailure,
  author  = {Temple, Simon W. P. and Rowbottom, Carl G.},
  title   = {Gross failure rates and failure modes for a commercial {AI}-based auto-segmentation algorithm in head and neck cancer patients},
  journal = {Journal of Applied Clinical Medical Physics},
  year    = {2024},
  volume  = {25},
  number  = {6},
  pages   = {e14273},
  doi     = {10.1002/acm2.14273}
}

@article{maierhein2024metricsreloaded,
  author  = {Maier-Hein, Lena and Reinke, Annika and Godau, Patrick and others},
  title   = {Metrics reloaded: recommendations for image analysis validation},
  journal = {Nature Methods},
  year    = {2024},
  volume  = {21},
  number  = {2},
  pages   = {195--212},
  doi     = {10.1038/s41592-023-02151-z}
}

@article{isensee2021nnunet,
  author  = {Isensee, Fabian and Jaeger, Paul F. and Kohl, Simon A. A. and Petersen, Jens and Maier-Hein, Klaus H.},
  title   = {{nnU-Net}: a self-configuring method for deep learning-based biomedical image segmentation},
  journal = {Nature Methods},
  year    = {2021},
  volume  = {18},
  number  = {2},
  pages   = {203--211},
  doi     = {10.1038/s41592-020-01008-z}
}

@article{babier2021openkbp,
  author  = {Babier, Aaron and Zhang, Binghao and Mahmood, Rafid and
             Moore, Kevin L. and Purdie, Thomas G. and McNiven, Andrea L. and
             Chan, Timothy C. Y.},
  title   = {{OpenKBP}: The Open-Access Knowledge-Based Planning Grand
             Challenge and Dataset},
  journal = {Medical Physics},
  year    = {2021},
  volume  = {48},
  number  = {9},
  pages   = {5549--5561},
  doi     = {10.1002/mp.14845}
}
\end{document}